\documentclass[11pt]{article}
\usepackage[colorlinks=true, citecolor=blue, urlcolor=blue, linkcolor=blue, breaklinks=true, pdfpagelabels=false]{hyperref}

\usepackage{hyperref}

\usepackage{amsmath,amsfonts}
\usepackage{graphicx}
\usepackage{epsfig,epsf}
\usepackage{bm}

\usepackage{dcolumn,subfig}

\def\a {\alpha}
\def\b {\beta}

\def\L {\Lambda}

\def\bar {\overline}

\def\p {\partial}

\def\be {\begin{equation}}
\def\ee {\end{equation}}
\def\beq {\begin{equation}}
\def\eeq {\end{equation}}
\def\bea {\begin{eqnarray}}
\def\eea {\end{eqnarray}}

\def\beq{\begin{equation}}
\def\eeq{\end{equation}}
\def\barr{\begin{array}}
\def\earr{\end{array}}

\def\opcit(#1){ {\em op. cit.}, #1}

\def\equationautorefname~#1\null{Eq.\,(#1)\null}
\def\pageautorefname\nobreakspace{p.}

\makeatletter\renewcommand{\p@subsection}{\thesection.}\makeatother%
 
\begin{document}

\renewcommand*{\thefootnote}{\fnsymbol{footnote}}


\begin{center}
{\Large\bf{Scattering unitarity with effective dimension-6 operators}}


\vspace{5mm}

{\bf Swagata Ghosh},$^{a,}$\footnote{swgtghsh54@gmail.com}
{\bf Rashidul Islam},$^{a,b,}$\footnote{islam.rashid@gmail.com}
and
{\bf Anirban Kundu}$^{a,}$\footnote{anirban.kundu.cu@gmail.com}

\vspace{3mm}
 ${}^a$ \ \ {\em{Department of Physics, University of Calcutta, \\
92 Acharya Prafulla Chandra Road, Kolkata 700009, India
}}

${}^b$ {\em{Department of Physics, Indian Institute of Technology, Guwahati 781039, India}}

\end{center}

\begin{abstract}

The effects of physics beyond the Standard Model may be parametrized by a set of higher-dimensional 
operators leading to an effective theory. The introduction of these operators makes the theory 
nonrenormalizable, and one may reasonably expect a violation of unitarity in $2\to 2$ scattering processes,
depending on the values of the Wilson coefficients of the higher dimensional operators. 
Bounds on these coefficients may be obtained from demanding that there be no such unitarity 
violation below the scale of the effective theory. We show, at the lowest level, how the new operators 
affect the scattering 
amplitudes with longitudinal gauge bosons, scalars, and $t\bar{t}$ in the final state, and find that one may 
expect a violation of unitarity even at the LHC energies with small values of some of the new Wilson
coefficients. For most of the others, such a violation needs large coefficients, indicating nonperturbative 
physics for the ultraviolet-complete theory, 
although a proper treatment necessitates the inclusion of even higher-dimensional operators. 
However, deviations from the Standard Model expectations may be 
observed with even smaller values for these coefficients. We find that $WW \to WW$, $WW\to ZZ$, 
and $ZZ\to hh$ scatterings are the best possible channels to probe unitarity violations.
\end{abstract}


\setcounter{footnote}{0}
\renewcommand*{\thefootnote}{\arabic{footnote}}


\section{Introduction}
\label{intro}

Even after a few years of running at the energy frontier, the Large Hadron Collider (LHC) is yet to 
give us any direct evidence of new physics (NP) beyond the Standard Model (SM), 
except for occasional will-o'-the-wisp signals 
drifting in and flickering out of existence. On the other hand, we have more than enough reasons to believe 
that NP exists in some form or other. This leads us to parametrize the NP in a model-independent way 
in terms of effective higher-dimensional operators, assuming all new degrees of freedom to be sufficiently heavy.
A well-used tool for low-energy physics, like Chiral Perturbation Theory or Heavy Quark Effective Theory,
this has also become a powerful weapon at the LHC energies. Use of such
effective field theories (EFT) to parametrize 
NP effects was first demonstrated in the seminal work of Ref.\ \cite{buchmuller}. 
While such theories are in general not renormalizable and hence cannot possibly be the ultimate 
ultraviolet-complete theory \footnote{The dimension-5 EFTs 
are renormalizable, this being a very special case. For $d\geq 6$ 
operators, the divergence coming from the loops can be absorbed only by even higher-dimensional 
operators \cite{1305.0017}, spoiling the renormalizability.}, higher dimensional operators can generate 
new tree-level interactions, which might include the effects of some hitherto unknown heavy degrees 
of freedom. 

It is important to have a complete basis of gauge and Lorentz-invariant operators 
of any given dimension; fortunately, this is well-known for the SM ({\em i.e.}, when all 
NP fields are integrated out) 
\footnote{For EFTs with low-mass NP fields, see, {\em e.g.}, 
Refs.\ \cite{1510.03443,1604.07365,1602.02645}.}. 
A minimal set of 59 dimension-6 operators was given in Ref.\ 
\cite{1008.4884} and later confirmed in other works \cite{1307.0478,1406.6376}. 
The importance of identifying and choosing a proper basis for the effective operators has 
recently been emphasized in Ref.\ \cite{1610.09618}.
 Refs.\ \cite{1412.1837,1603.03660} show a practical way to 
construct and use the SM EFTs.
However, only a subset of the higher-dimensional operators is relevant to study a particular process. 
For example, effects of the new operators on Higgs physics were studied, with a subset of all 
the dimension-6 operators, in Refs.\ 
\cite{0703164,0907.5413,1302.5661,1303.1812,1303.3876,1308.2255}. 
Experiments can constrain the Wilson coefficients of the higher dimensional operators,
either directly or indirectly
\cite{1404.3667,1404.5343,1406.7320,1311.3107,1411.0669,1508.05060,reuter}. 
Among other interesting uses of the EFT formalism, one can mention the attempt to address the naturalness 
issue of the Higgs boson mass \cite{Barshalom-Soni-Wudka}. 

There are many equivalent bases to write a complete set of operators ({\em e.g.}, see 
\cite{1504.01707}). We will use the Strongly Interacting Light Higgs (SILH) basis, 
as shown in \cite{GGPR,contino}, 
but work with only those operators that are relevant for the scattering processes. 
As our aim is to look at possible unitarity violations in $2\to 2$ scattering 
processes when such higher dimensional operators are present, we would prefer a basis that 
couples the Higgs sector strongly with the NP \footnote{Any other basis could have been used too. The choice 
of the most effective basis will also be governed by the observables the experiments measure. However,
one has to be careful of not introducing redundant operators.}. 
Let $\L$ be the scale where the new degrees of freedom 
show up, so this will act as the cutoff for the effective theory. We will keep $\Lambda$ a free 
parameter, as the Wilson coefficients (WC) scale trivially with $\L^2$. We emphasize that setting $\L = 1$ TeV 
as a {\em fiducial mark} does not necessarily mean that one must observe NP effects beyond that, say 
at the LHC. 

The four-point vertices coming from the dimension-6 operators generally contain a prefactor 
of $c_i v^2/\Lambda^2$, where $c_i$ is the respective WC, and $v$ is the vacuum expectation value (VEV)
of the Higgs field. Thus, the cancellation of the bad high-energy 
behaviour is affected. It is affected even more if the prefactor contains momentum dependence. 
However, we will always keep terms of the order of $1/\Lambda^2$. Going beyond that would necessitate 
the consideration of even higher-dimensional operators. This is to be kept in mind as the new operators 
will manifest themselves in two different ways, which we discuss below.

The scattering of longitudinally polarized gauge bosons to two-particle final states gave, perhaps, the 
strongest motivation to have a Higgs sector in the SM \cite{LQT}, because without the Higgs field the scattering 
amplitudes tend to violate unitarity at high enough energies. In the SM, the bad high-energy behaviour of the 
scattering amplitudes is completely tamed because of the precise gauge structure and the Higgs mechanism. 
With effective operators, one may re-introduce the bad behaviour because such cancellations no longer hold, 
unless the cutoff scale $\Lambda$ is extremely large. This puts a bound on the effective WCs, originating from 
the fact that the magnitude of real part of any partial wave amplitude must be less than 
$\frac12$. A study somewhat similar in approach, assuming that the SM vertices may allow some deviations from 
their predicted values, through possibly higher-order interactions, was undertaken in Refs.\ \cite{cik,rashidul2}.  

The new dim-6 operators contribute in two ways to $2\to 2$ scattering processes. First, they contribute to the 
scattering vertices in a non-trivial way, whose examples we will give later. They can change the vertex factors 
by terms typically going as $v^2/\Lambda^2$. They can also introduce momentum dependence to the erstwhile 
momentum-independent vertex factors. The second point is not very obvious 
and is often neglected in the literature. The new operators can also modify the kinetic terms in the Lagrangian; 
for example, an operator of the form $(c/\L^2) W^{\mu\nu}W_{\mu\nu} \Phi^\dag\Phi$, where $\Phi$ is the SM
scalar doublet, 
produces an extra contribution of $cv^2/2\L^2$ to the kinetic term. If $v\ll \L$, this correction is negligible, 
but in principle this affects the scattering amplitudes by redefining the fields, as well as modifying the
corresponding vertex factors for the relevant Feynman diagrams. The Higgs VEV gets modified, and so 
do the masses of the gauge bosons and fermions where the Higgs VEV is fed. 
The WCs of some of the field-redefining 
operators, as we will see, can be tightly constrained from electroweak precision observables. 
A similar study in another choice of basis was performed in Ref.\ \cite{1411.5026}, but without the 
proper normalization of the kinetic terms. 

The effects of NP on the partial wave amplitudes are expected to be small, suppressed typically by 
$v^2/\L^2$ unless there is some momentum-dependent enhancement.
However, $\L$ may very well be within a few TeV, as predicted by most of the NP theories. 
In fact, if one takes the cutoff scale to be $\Lambda \sim {\cal O}(1)$ TeV, even at the LHC energies 
one may observe violations of the unitarity bound with small to moderate WC. 
To get a meaningful estimate, one must not go beyond the small-WC limit, if $d > 6$ 
operators are not taken into account. 

In this paper, we will try to see how the scattering amplitudes (the zero-th partial wave, to be more precise) 
behave with the introduction of these new operators, and when we may expect a violation of unitarity. 
As expected, if $\Lambda\to\infty$, all the amplitudes will be well-behaved. 
We will show that if the NP is indeed at the electroweak scale, one may expect, at the LHC, 
a violation of unitarity even with perturbative new couplings. A similar study will be applicable to the 
future $e^+e^-$ colliders too, and in a much cleaner environment. Effects of the dimension-6 operators 
in hadronic \cite{1511.05170,1702.05753} and leptonic colliders \cite{ellis,hamzeh} 
have already been discussed in the literature, and we will comment on their bounds later. 
Another interesting point 
from the collider perspective that we will not discuss is the introduction of new three- and four-point 
interactions from the effective operators, as shown in the Appendix. They can in principle affect 
processes like Higgs production from gluon fusion, or its decay to a photon and a $Z$. 

In Section II, we enlist the set of operators that might be interesting to study 
unitarity violation. We also illustrate how one properly normalizes the fields. We discuss two sets of 
operators, bosonic and fermionic; the former involves only bosonic fields and the latter involves fermionic 
fields too. While the bosonic operators can contribute even to $V_L V_L \to t\bar{t}$ scattering (where $V_L$ is
any generic longitudinally polarized gauge boson) amplitudes, fermionic operators can never contribute 
to bosonic final states. 
The bounds on the corresponding WCs are shown in Section III. 
As expected, $W_LW_L \to W_L W_L$, $W_LW_L \to Z_LZ_L$, and 
$Z_LZ_L \to hh$  scatterings almost invariably put the strongest bound to 
whichever operators it gets a contribution from. We summarize and conclude in Section IV, and relegate 
some detailed calculation, including that of the modified vertex factors, to the Appendix.

\section{Formalism}
\subsection{The effective Lagrangian}

For any scattering we can decompose the amplitude into partial waves
\be
A = 16\pi \sum_{\ell=0}^\infty \, (2\ell+1) \, P_\ell(\cos\theta)\, a_\ell,
\ee
and by virtue of the optical theorem which relates the cross-section with the imaginary part 
of the amplitude for zero scattering angle, one gets \footnote{To treat all possible spins of incoming 
and outgoing particles, one should use Wigner's $D$-functions, but $D^0_{00}$ is directly
related with $a_0$. See Ref.\ \cite{rashidul2} for details.}
\be
|a_\ell|^2 = {\rm Im} \, a_\ell \, \, \Rightarrow\, \,  {\rm Re}\, a_\ell \leq \frac12\,.
\ee
We will be interested in $\ell=0$ partial waves only. 

We start with the set of SILH operators as given in Ref.\ \cite{contino}, and follow their notation 
and convention. In particular, $\bar c_i$ denotes any generic WC, and the operator that comes with 
$\bar c_i/\L^2$ is denoted by $O_i$. Let us first pick up only 
those operators that lead to $2\to 2$ bosonic scatterings, and set the generic cut-off scale at some 
high momentum $\Lambda$. Our fiducial marker is at $\Lambda = 1$ TeV but the WCs
scale with $\Lambda^2$, so the actual bound on any generic WC should be read as 
$\bar{c}_i (\Lambda/1~{\rm TeV})^2$. We will show our
results for $\sqrt{s}=2$ TeV, the typical parton-level energy at the LHC, but this {\em does not mean} that 
we are necessarily in the new physics regime, as already emphasized. The values of $\bar c_i$ for
$\L=1$ TeV are denoted by $C_i$. 
We confine ourselves to $|C_i| < 1$ so that $d>6$ operators can be neglected.

For bosonic scatterings, the relevant terms are
\bea
 {\cal L}_{\rm boson} = \frac{1}{\Lambda^2}\sum_i \bar c_i O_i 
 &=&
   \dfrac{\bar c_H}{\Lambda^2} \ \p^\mu(\Phi^\dag \Phi) \ \p_\mu(\Phi^\dag \Phi)
 + \dfrac{\bar c_T}{\Lambda^2} \ ( \Phi^\dag \overleftrightarrow D^\mu \Phi )( \Phi^\dag \overleftrightarrow D^\mu \Phi )
 - \dfrac{\bar c_6 \lambda}{\Lambda^2} \ (\Phi^\dag \Phi)^3
 \nonumber\\
 &&
 + \dfrac{i g \bar c_W}{2\Lambda^2}
    \ ( \Phi^\dag \tau^i \overleftrightarrow D^\mu \Phi )
     \ ( D^\nu W_{\mu\nu} )^i
 + \dfrac{ig' \bar c_B}{2\Lambda^2}
    \ ( \Phi^\dag \overleftrightarrow D^\mu \Phi )
     \ ( \p^\nu B_{\mu\nu} )
 \nonumber\\
 &&
 + \dfrac{ig\bar c_{HW}}{2\Lambda^2}
    \ (D^\mu \Phi)^\dag \tau^i (D^\nu \Phi)
     \ W^i_{\mu\nu}
 + \dfrac{ig'\bar c_{HB}}{2\Lambda^2}
    \ (D^\mu \Phi)^\dag (D^\nu \Phi) \ B_{\mu\nu}
 \nonumber\\
 &&
 + \dfrac{{g'}^2 \bar c_\gamma}{\Lambda^2} \ (\Phi^\dag \Phi) \ B_{\mu\nu} B^{\mu\nu}
 + \dfrac{g_s^2 \bar c_g}{\Lambda^2} \ (\Phi^\dag \Phi) \ G^a_{\mu\nu} G^{a\mu\nu}\nonumber\\
 &&
 + \dfrac{g^3 \bar c_{3W}}{\Lambda^2} \epsilon_{ijk} 
 W^{i\nu}_\mu W^{j\a}_\nu W^{k\mu}_{\a}\,,
\label{eqlboson}
\eea
where $\tau_i$s are the Pauli matrices, 
$g'$, $g$, and $g_s$ are the U(1)$_Y$, SU(2)$_L$, and SU(3)$_c$ gauge couplings
respectively, and $\Phi^\dag \overleftrightarrow D^\mu \Phi = \Phi^\dag D^\mu\Phi - 
(D^\mu\Phi)^\dag\Phi$. The gluon operator $O_g$ will not be relevant for us as we do not consider 
final states involving gluons, but this will lead to a direct production of the Higgs boson from 
gluon fusion. These operators can contribute to $W_LW_L (Z_L Z_L) \to 
t\bar{t}$ too, with the fermionic vertex being SM-like and the bosonic vertex involving the 
contributions from the dimension-6 operators\footnote{Only top quark, because of its mass, can
contribute to the $\ell=0$ channel. The other fermionic final states occur at higher $\ell$ and 
therefore give a much weaker constraint \cite{1411.5026}. That is why we do not consider $q\bar{q}\to VV$ 
scattering where the $q$ and $\bar{q}$ come from the initial protons. This is also true for the
fermionic dimension-6 operators.}
Note that there are other operators involving three
gauge tensors but the last one, $O_{3W}$, is the only one that contributes \cite{1411.5026}. 
At the same time, $O_{3W}$ is not produced at the tree-level \cite{0703164} and hence $\bar c_{3W}$ 
is expected to have a further loop suppression. 

The relevant fermionic operators are
\bea
{\cal L}_{\rm fermion} = \frac{1}{\Lambda^2}\sum_j \bar c_j O_j&=&
 \left( 
 \frac{\bar{c}_u}{\L^2} \, y_u \bar{Q}_L u_R\, \Phi^c \Phi^\dag\Phi  
 + \frac{i\bar{c}_{Hud}}{\L^2} \, (\bar u_R \gamma^\mu d_R)\, (\Phi^{c\dag} \overleftrightarrow 
 D_\mu \Phi) +{\rm h.c.} 
 \right) \nonumber\\
 &&+ \frac{i\bar{c}_{Hq}}{\L^2} \, (\bar Q_L \gamma^\mu Q_L)\, (\Phi^\dag \overleftrightarrow D_\mu \Phi) 
 + \frac{i\bar{c}'_{Hq}}{\L^2}\, (\bar Q_L \tau^i \gamma^\mu Q_L)\, (\Phi^\dag \tau^i 
 \overleftrightarrow D_\mu \Phi)\nonumber\\
 && + \frac{i\bar{c}_{Hu}}{\L^2}\, (\bar u_R \gamma^\mu u_R)\, (\Phi^\dag \overleftrightarrow D_\mu \Phi) 
 + \frac{i\bar{c}'_{Hd}}{\L^2} \, (\bar d_R \tau^i \gamma^\mu d_R)\, (\Phi^\dag \tau^i 
 \overleftrightarrow D_\mu \Phi)\nonumber\\ 
  && + \frac{g' \bar{c}_{uB}}{\L^2} y_u\, \bar Q_L \Phi^c \sigma_{\mu\nu} u_R\, B^{\mu\nu} 
 + \frac{g \bar{c}_{uW}}{\L^2} y_u\, \bar Q_L \tau^i \Phi^c \sigma_{\mu\nu} u_R\, W^{i\mu\nu} \nonumber\\
 && + \frac{g' \bar{c}_{dB}}{\L^2} y_d\, \bar Q_L \Phi \sigma_{\mu\nu} d_R\, B^{\mu\nu} 
 + \frac{g \bar{c}_{dW}}{\L^2} y_d\, \bar Q_L \tau^i \Phi \sigma_{\mu\nu} d_R\, W^{i\mu\nu}\,,
 \label{eqlfermion}
 \eea 
where we have neglected operators involving lepton and gluon fields. The Feynman rules involving 
the new operators are shown in Appendix A, while the detailed structure of one of the operators is 
shown in Appendix B. 

\subsection{Field redefinition}

Let us spend some time on the field redefinition here, to bring the kinetic terms to 
their canonical form. From the expression of $O_W$ in Eq.\ (\ref{OW3}), one finds that 
the following terms
\bea
 O_W
 &\supset&
 \dfrac{g^2}{4}\, v^2
 \left[
   (\p^\mu W^{+\nu} - \p^\nu W^{+\mu}) (\p_\mu W^-_\nu - \p_\nu W^-_\mu)
 \right]
 \nonumber\\
 &&
 + \dfrac{g^2}{4 \cos\theta_W} v^2
 \left[
  - \cos\theta_W \p^\nu Z^\mu (\p_\mu Z_\nu - \p_\nu Z_\mu)
  - \sin\theta_W \p^\nu Z^\mu (\p_\mu A_\nu - \p_\nu A_\mu)
 \right]
\eea
contribute to the 2-point functions. 
Similar contributions come from other operators also. All the contributions to
the 2-point functions from the operators can be summed up as
\bea
 {\cal L}_{\rm 2point}&=&
   \frac{\bar c_H v^2}{\Lambda^2}\, (\p_\mu h) (\p^\mu h)
 + g^2_s v^2 \frac{\bar c_G}{2 \Lambda^2}\, G^a_{\mu\nu} G^{a\mu\nu}
 + g^2 v^2 \frac{\bar c_W}{4 \Lambda^2} W^-_{\mu\nu} W^{+\mu\nu}
  + {g'}^{2} \sin^2\theta_W v^2 \frac{\bar c_\gamma}{2 \Lambda^2} Z_{\mu\nu} Z^{\mu\nu}
 \nonumber\\
 &&
 + {g'}^2 v^2 \frac{\bar c_B}{8 \Lambda^2} Z_{\mu\nu} Z^{\mu\nu}
 + g^2 v^2 \frac{\bar c_W}{8 \Lambda^2} Z_{\mu\nu} Z^{\mu\nu}
  + g^2 \sin^2\theta_W v^2 \frac{\bar c_\gamma}{2 \Lambda^2} A_{\mu\nu} A^{\mu\nu}
 \nonumber\\
 &&
 - g g' \sin^2\theta_W v^2 \frac{\bar c_\gamma}{\Lambda^2} A_{\mu\nu} Z^{\mu\nu}
 - g g' v^2 \frac{\bar c_B}{8 \Lambda^2} A_{\mu\nu} Z^{\mu\nu}
 + g g' v^2 \frac{\bar c_W}{8 \Lambda^2} A_{\mu\nu} Z^{\mu\nu}\,.
 \label{kin_ops}
\eea

With this, one should add the canonical kinetic terms, which gives
\be
{\cal L}_{\rm kinetic} =
   \frac12 (\p_\mu \bar h) (\p^\mu \bar h)
 - \frac14 \bar G^a_{\mu\nu} \bar G^{a\mu\nu}
 - \frac12 \bar W^-_{\mu\nu} \bar W^{+\mu\nu}
 - \frac14 \bar Z_{\mu\nu} \bar Z^{\mu\nu}
 - \frac14 \bar A_{\mu\nu} \bar A^{\mu\nu}\,
\ee
where
\bea
 \bar h &=& \sqrt{ 1 + 2 \frac{\bar c_H v^2}{\Lambda^2} } \ h \equiv \sqrt{N_h}\, h\,,\nonumber\\
 \bar G^a_\mu &=& \sqrt{ 1 - 2 g^2_s v^2 \frac{\bar c_G}{\Lambda^2} } \ G^a_\mu 
 \equiv \sqrt{N_G}\, G^a_\mu\,,\nonumber\\
 \bar W^\pm_\mu &=& \sqrt{ 1 - g^2 v^2 \frac{\bar c_W}{2 \Lambda^2} } \ W^\pm_\mu
 \equiv \sqrt{N_W}\, W^\pm_\mu\,, \nonumber\\
 \bar Z_\mu &=& \sqrt{ 1 - 2 g^{\prime2} \sin^2\theta_W v^2 \frac{\bar c_\gamma}{\Lambda^2}
 - {g'}^2 v^2 \frac{\bar c_B}{2 \Lambda^2}
 - g^2 v^2 \frac{\bar c_W}{2 \Lambda^2} } \ Z_\mu \equiv \sqrt{N_Z}\, Z_\mu\,,\nonumber\\
 \bar A_\mu &=& \left[ 1 - g^2 \sin^2\theta_W v^2 \frac{\bar c_\gamma}{\Lambda^2} \right] A_\mu
 + \left[ 2 g g' \sin^2\theta_W v^2 \frac{\bar c_\gamma}{\Lambda^2}
 + g g' v^2 \frac{\bar c_B}{4 \Lambda^2}
 - g g' v^2 \frac{\bar c_W}{4 \Lambda^2} \right] \ Z_\mu \nonumber\\
 &\equiv& N_A\, A_\mu + N_{AZ}\, Z_\mu\,.
 \label{field01}
\eea
This gives the field redefinitions; also, this shows that one may not extend the $\bar{c}_i$s beyond their 
range of validity, given by
\be
\bar c_W < 83\,\left(\frac{\Lambda}{1~{\rm TeV}}\right)^2\,,\ \ 
\bar c_B < 276\,\left(\frac{\Lambda}{1~{\rm TeV}}\right)^2\,,\ \ 
\bar c_\gamma < 179\left(\frac{\Lambda}{1~{\rm TeV}}\right)^2\,.
\ee
This, again, naively assumes that there are no other operators of mass dimension greater than 
6, and therefore one does not need to take these constraints too seriously.
Stronger constraints come from electroweak precision observables, like the $\rho$-parameter, 
as we will soon show. 

If $\bar{c}_i v^2/\Lambda^2 \ll 1$, one can invert these relations 
by a binomial expansion and obtain
\bea
 h &\to & h
 \left[
   1 - \dfrac{\bar c_H}{\Lambda^2} v^2
 \right]\,,\nonumber\\
 G^a_\mu &\to & G^a_\mu
 \left[
   1 + \dfrac{\bar c_G}{\Lambda^2} g^2_s v^2
 \right]\,,\nonumber\\
 W^\pm_\mu &\to& W^\pm_\mu
 \left[
   1 + \dfrac{\bar c_W}{\Lambda^2} \dfrac{g^2}{4} v^2
 \right]\,, \nonumber\\
 Z_\mu &\to& Z_\mu
 \left[
   1 + \dfrac{\bar c_\gamma}{\Lambda^2} {g'}^2 \sin^2\theta_W v^2
   + \dfrac{\bar c_W}{\Lambda^2} \dfrac{g^2}{4} v^2
   + \dfrac{\bar c_B}{\Lambda^2} \dfrac{g^{\prime2}}{4} v^2
 \right]\,,\nonumber\\ 
 A_\mu &\to& A_\mu
 \left[
   1 + \dfrac{\bar c_\gamma}{\Lambda^2} g^2 \sin^2\theta_W v^2
 \right]
 + Z_\mu
   \dfrac{\bar c_W - \bar c_B - 8 \bar c_\gamma \sin^2\theta_W}{\Lambda^2}
   \dfrac{g g'}{4} v^2\,.
\eea
Note that the binomial expansion is valid only in the proper limit. For numerical 
evaluations, we work with the exact definitions. However, the vertex factors 
of the effective theory depend on the WCs, and in the limit when 
binomial expansion fails, they become non-perturbative and 
so large as to make the higher-order effects more important 
than the tree-level ones. Fortunately, the possible unitarity violations, in at least one channel,
occur much before that range.

The expressions for the particle masses, except for the photon which remains massless because of unbroken 
electromagnetism, are also modified 
and can be read off from the bilinear term. The couplings now depend on the higher-dimensional
WCs, so that the particle masses are tuned to their experimental values: 
\bea
 m^2_h &=&
 \left[1 - 2 v^2 \dfrac{\bar c_H}{\Lambda^2}\right]
 (3 \lambda v^2 - \mu^2)
 + \dfrac{15}{4} \lambda v^4 \dfrac{\bar c_6}{\Lambda^2}\,,\nonumber\\
 m^2_W &=&
 \dfrac{g^2 {v'}^2}{4}
 \left[
 1 + g^2 {v'}^2 \dfrac{\bar c_W}{2 \Lambda^2}
 \right]\,,\nonumber\\
 m^2_Z &=&
 \dfrac{g^2 {v'}^2}{4 \cos^2\theta_W}
 \left[
 1 + g^2 {v'}^2 \dfrac{\bar c_W}{2 \Lambda^2} + {g'}^2 {v'}^2 \dfrac{\bar c_B}{2 \Lambda^2}
 + 2 {g'}^2 {v'}^2 \sin^2\theta_W \dfrac{\bar c_\gamma}{\Lambda^2}
 - 2 {v'}^2 \dfrac{\bar c_T}{\Lambda^2}
 \right]\,.
\eea
Here, $v'$ is the modified Higgs VEV which follows from the redefinition 
$m_h^2 = 3\lambda {v'}^2 - {\mu'}^2$, where $v'$ and $\mu'$ contain the 
effects of the operators $O_6$ and $O_H$. 
Note that $O_W$ alone does not affect the tree-level $\rho$ parameter. The other operators do, and from
$T-T_{\rm SM}=0.08\pm 0.12$
where $\rho = 1 +\alpha T$, one gets 
\be
-0.44 < c_B < 0.08\,,\ \ \ 
-0.48 < c_\gamma < 0.09\,,\ \ \ 
-2.5\times 10^{-3} < c_T < 0.013\,,
\ee
where $c_i = \bar{c}_i (\Lambda / 1~{\rm TeV})^2$. While these bounds are somewhat stronger than the Froissart 
bounds for the WCs, the deviation of the cross-section from the SM expectations should be observable 
at about these values, or even less, as we will see later. This is why we do not talk about the precision 
observable bounds any further. 

One now has to write down the scattering amplitudes not only involving the dimension-6 terms of the 
effective Lagrangian, but also in terms of the normalized fields. This in turn means that even the 
SM vertices as well as the propagators are modified and become functions of the WCs. These vertex factors 
are enlisted in Appendix A. 
At the same time, we keep ourselves confined to such small values of $c_i$ that only the term
linear in $c_i/\L^2$ is sufficient. One may ask whether we need to take into account the field normalizations 
in that case. The answer is yes, as such lowest-order corrections appear even when the SM amplitude is 
calculated with the normalized fields. Sometimes the corrections coming from the vertices are cancelled or 
enhanced by a similar correction coming from the fields.

We will discuss only about those scatterings that can be observed either by the Large Hadron Collider (LHC) 
or the next generation International Linear Collider (ILC). They include $WW\to WW$, $WW\to ZZ$, $ZZ\to ZZ$, 
$WW \to hh$, $ZZ\to hh$, $WW \to t\bar{t}$, and $ZZ\to t\bar{t}$,
with crossed channels included wherever necessary, and the longitudinal mode is 
implied for the gauge bosons.
As we will see, only bosonic scatterings produce any useful constraints.

\section{Bounds on the Wilson coefficients}

 \begin{table}[htbp]
  \centering
   \begin{tabular}{|| c || c | c | c | c | c | c | c | c | c ||}
    \hline\hline
    PROCESS     & $O_6$   & $O_T$   & $O_H$   & $O_\gamma$  & $O_W$   & $O_{HW}$ & $O_B$   & $O_{HB}$ & $O_{3W}$ \\ \hline
    $WW\to WW$  &         &         & $\otimes$ & $\otimes$ & $\surd$ $\otimes$ & $\surd$  &  $\otimes$   & $\surd$ & $\surd$   \\ \hline
    $WW\to ZZ$  &         &  $\surd$  &  $\otimes$ & $\surd$ $\otimes$  & $\surd$ $\otimes$ & $\surd$  
    & $\otimes$  & $\surd$ & $\surd$  \\ \hline
    $ZZ\to ZZ$  &         & $\surd$ & $\otimes$ & $\surd$ $\otimes$ & $\otimes$ & $\surd$  & $\otimes$ 
    & $\surd$ &   \\ \hline
    $WW\to hh$  & $\surd$ &         & $\surd$ $\otimes$ &         & $\surd$ $\otimes$ 
    & $\surd$  &         &   $\surd$ & \\ \hline
    $ZZ\to hh$  & $\surd$ & $\surd$ & $\surd$ $\otimes$ & $\surd$ $\otimes$ & 
    $\surd$ $\otimes$ & $\surd$  & $\surd$ $\otimes$ & $\surd$  &  \\ \hline
    $WW \to t\bar{t}$ &      &            &  $\otimes$  & $\otimes$  &$\surd$ $\otimes$& $\surd$  & $\otimes$   &
     & $\surd$    \\ \hline
    $ZZ \to t\bar{t}$ &      & $\surd$  & $\otimes$  & $\surd$ $\otimes$ & $\otimes$ & $\surd$ & 
    $\otimes$ & $\surd$ & \\ \hline
   \end{tabular}
   \caption{\small{Dimension-6 operators affecting the bosonic and fermionic 
   scatterings. The entries marked with $\surd$ 
   are affected by the modification of the SM vertices. The entries marked with $\otimes$ are affected by the 
   wavefunction normalization.}} 
      \label{tab:total}
 \end{table}

\begin{figure}[htbp]
\begin{center} 
\includegraphics[width=8cm]{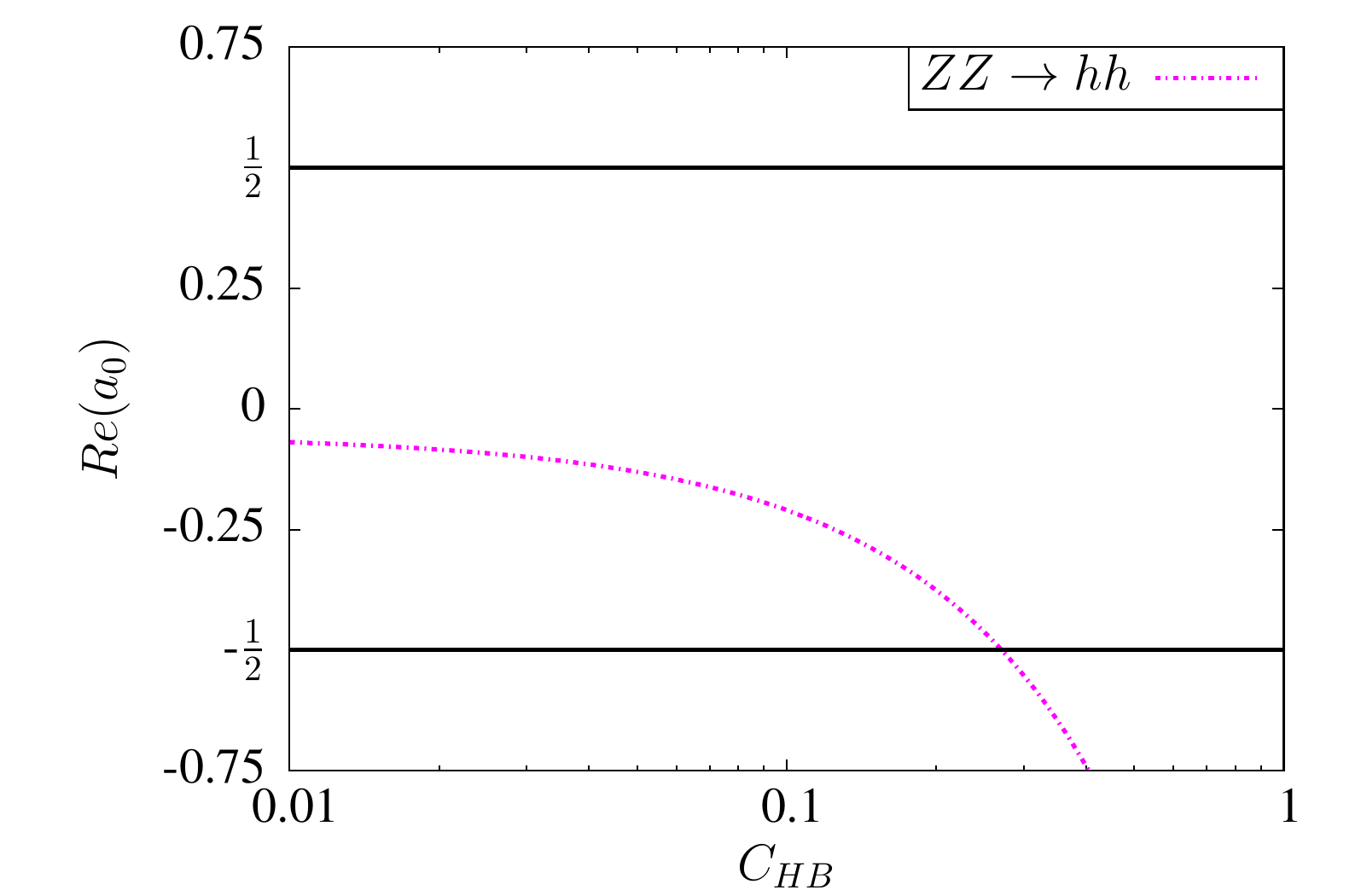}
\includegraphics[width=8cm]{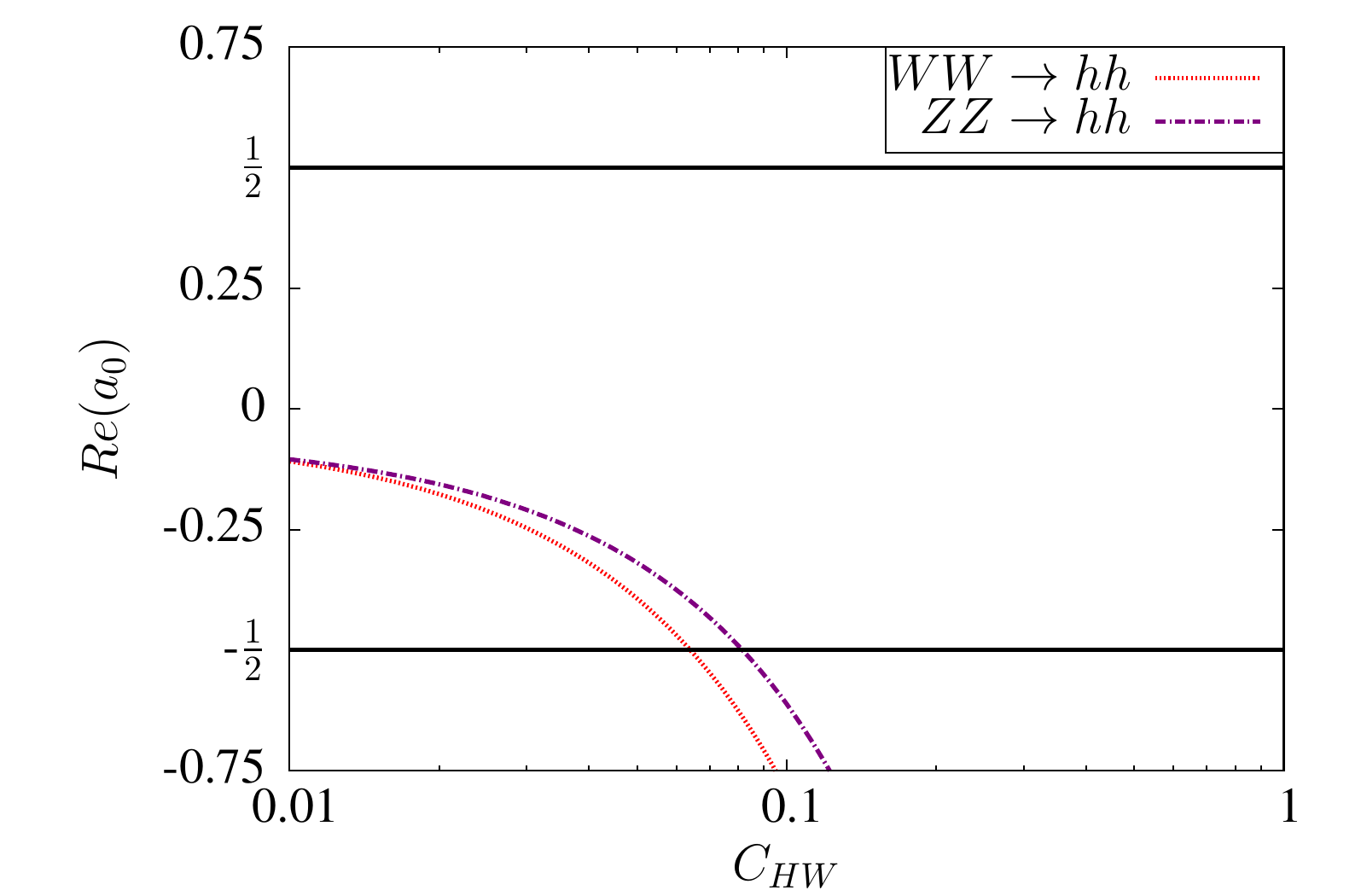}\\
\includegraphics[width=8cm]{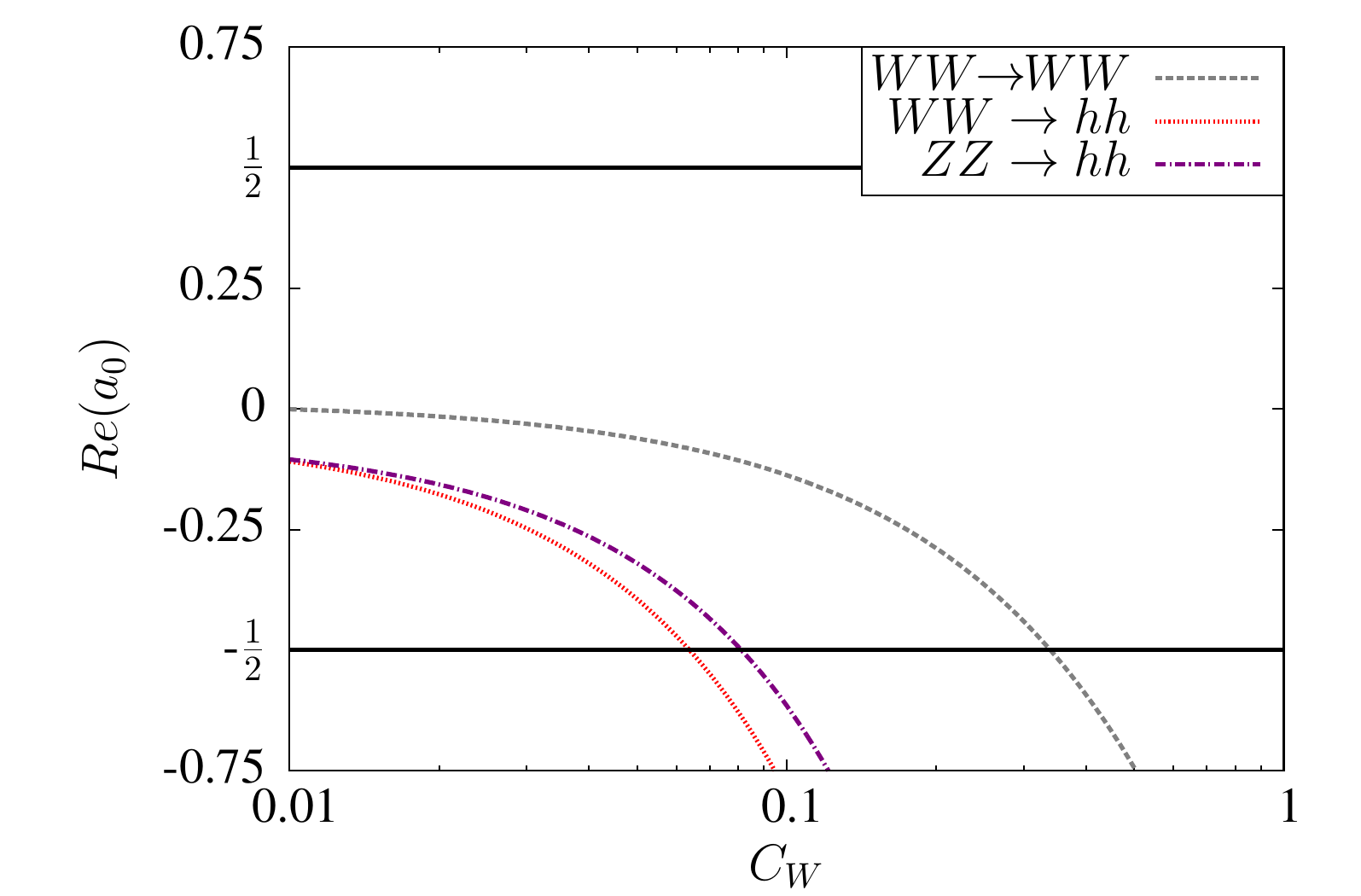}
\includegraphics[width=8cm]{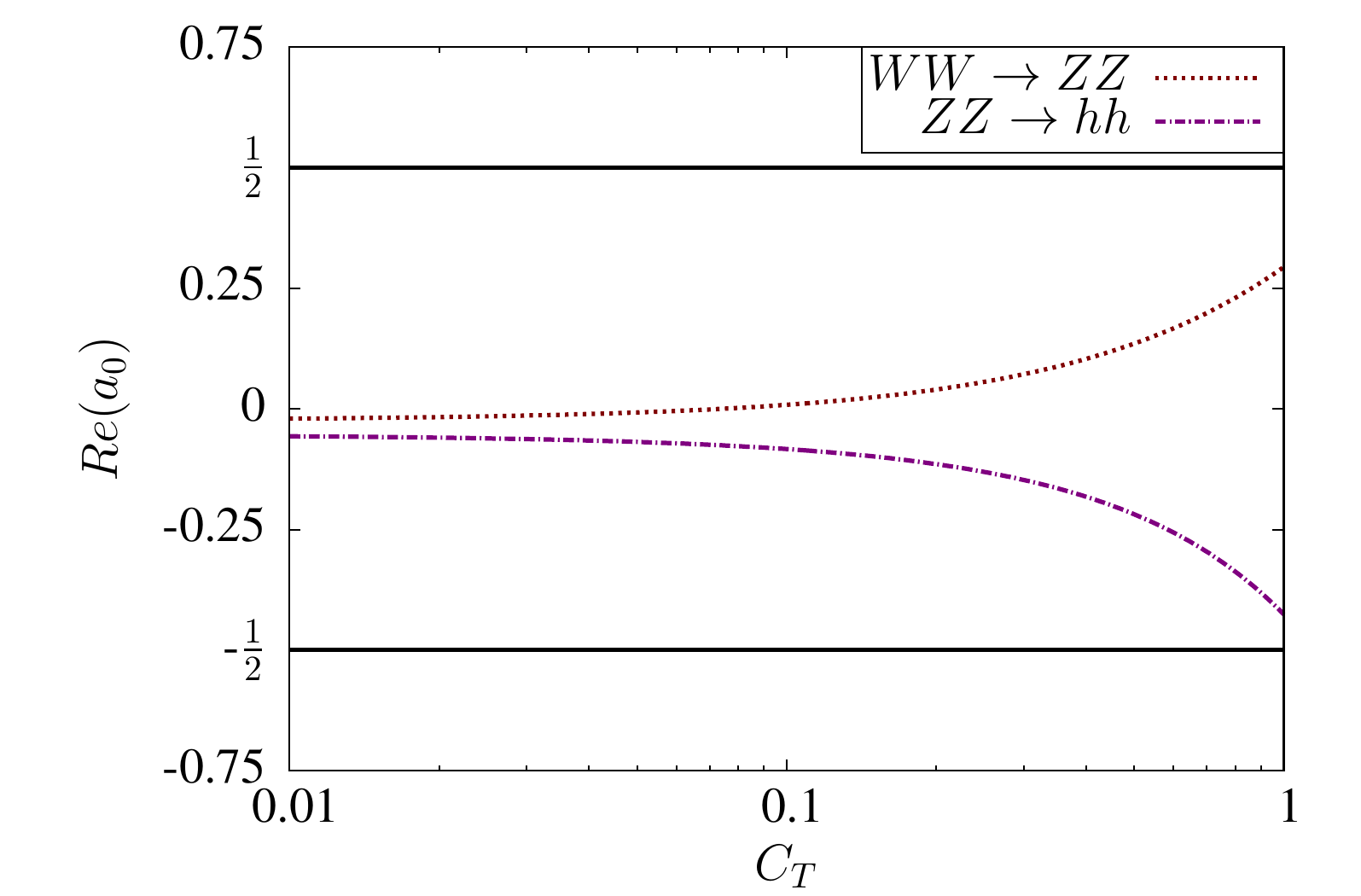}\\
\includegraphics[width=8cm]{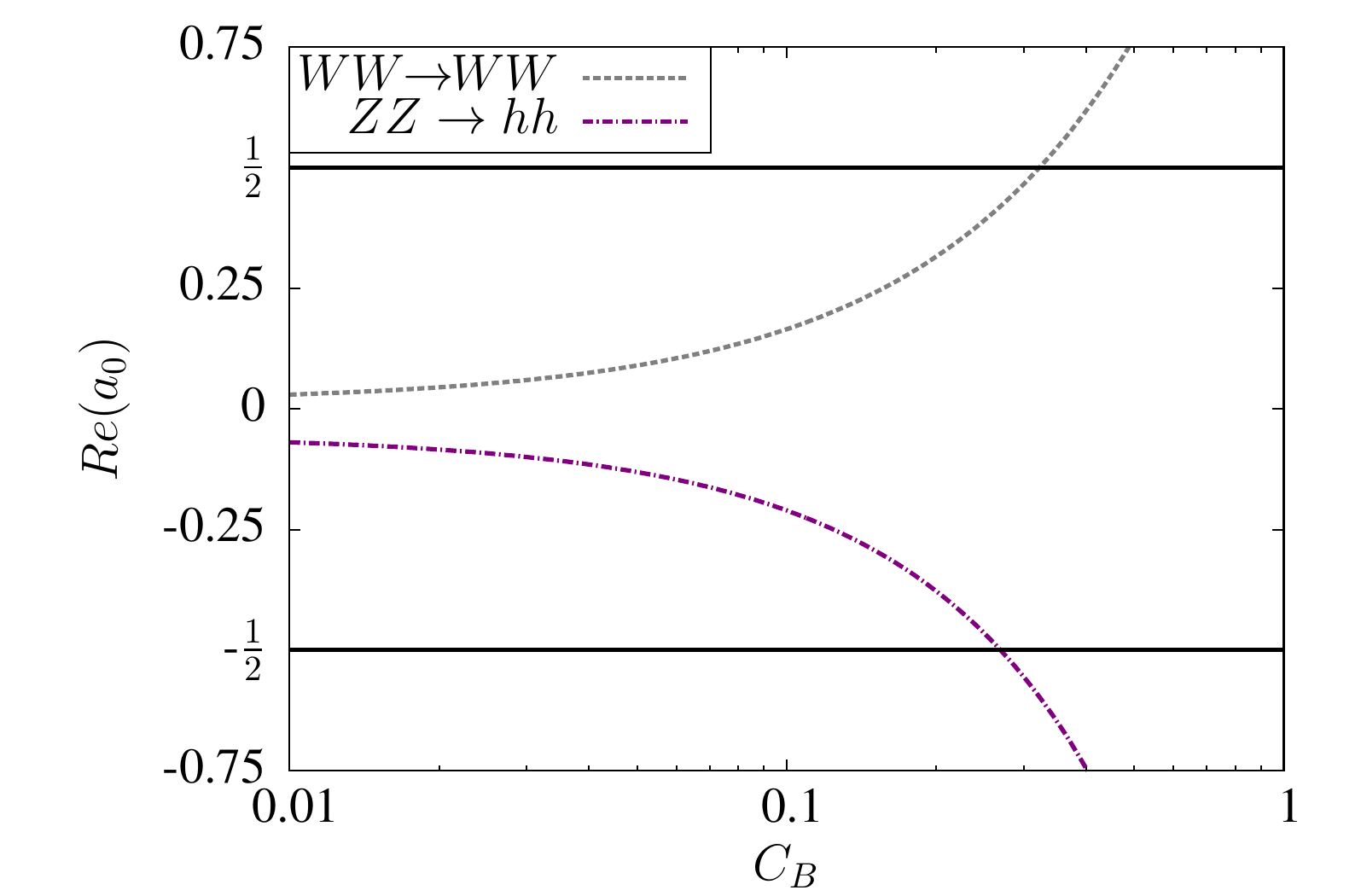}
\includegraphics[width=8cm]{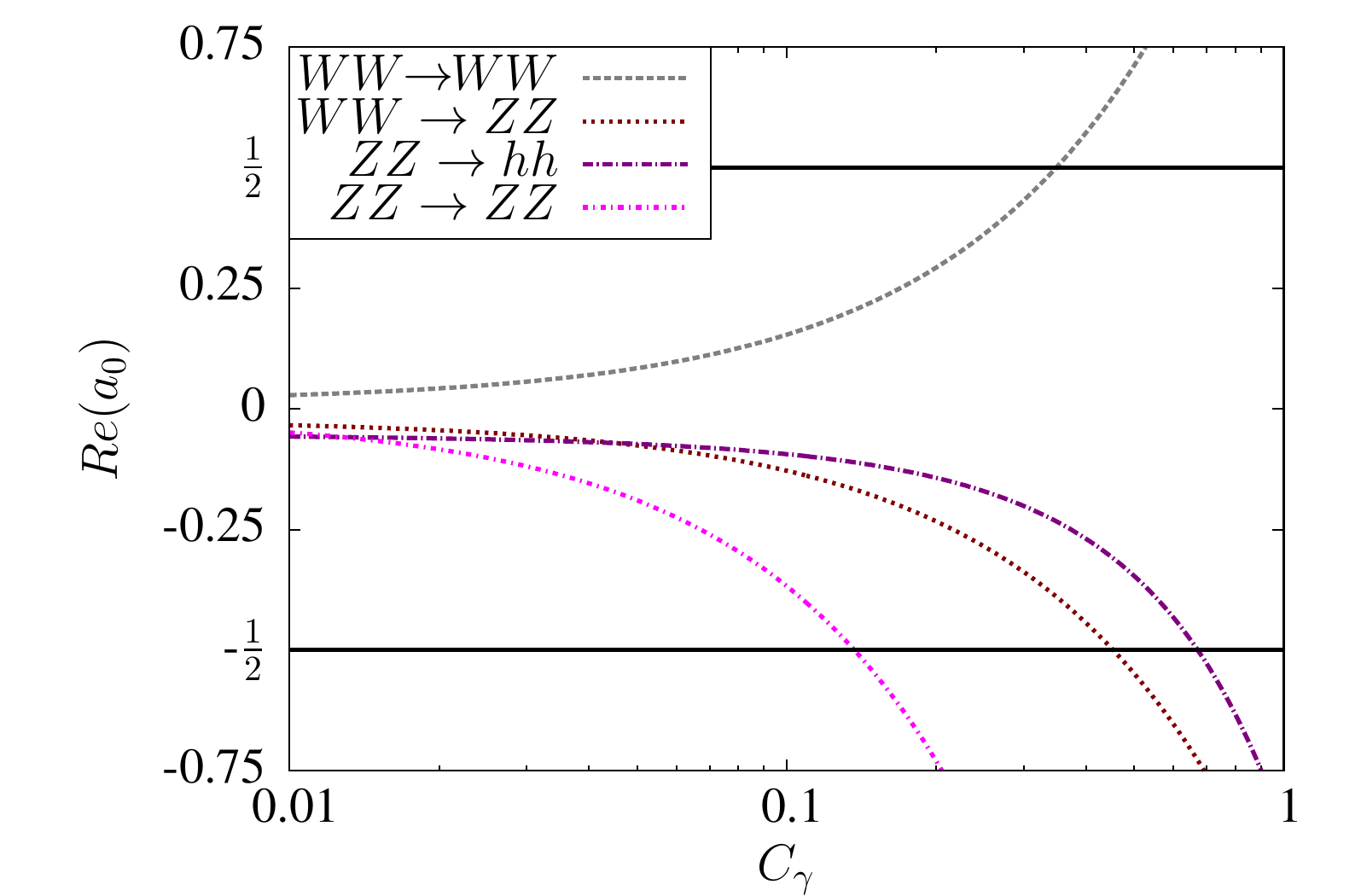}
\end{center} 
\caption{\small The unitarity limits on the effective Wilson coefficients, where they violate the 
bound $|{\rm Re} a_0|\leq \frac12$. We have taken $\sqrt{s}=2$ TeV and $\L = 1$ TeV, the coefficients 
scale with $\L^2$.  }
\label{fig:wcbound}
\end{figure}

To get the bounds on the WCs, we fix $\sqrt{s}=2$ TeV, which is a typical parton-level 
value for the proton-proton collision at the LHC with $\sqrt{s}=13$ or 14 TeV.
We use FeynArts/FormCalc \cite{feynarts, formcalc} 
to calculate the helicity amplitudes using the
FeynArts model files for the effective Lagrangian generated by FeynRules \cite{feynrules}.
This gives us all the field normalizations as well as the vertex factors. We then observe how 
the zero-th partial wave amplitude, $a_0$, varies with the WCs; the bound comes from
$|a_0| \leq \frac12$. For $WW (ZZ) \to t\bar{t}$, we use the helicity amplitude for 
$00\to ++$ as this gives the tightest constraints. 

In general, such operators also contribute to the higher-$\ell$ states. However, as has been shown
in Ref.\ \cite{1411.5026,rashidul2}, such constraints are always weaker than those coming from $\ell=0$. 
An intuitive way to understand this is that the worst high-energy behaviour of $\ell=0$ partial wave
goes as $s/m^2$ whereas it is $\sqrt{s}/m^2$ for $\ell=1$. We have found no $\ell=1$ constraints that are
either stronger than the same coming from $\ell=0$, or lie in the perturbative domain.

\begin{figure}[htbp]
\begin{center} 
\includegraphics[width=8cm]{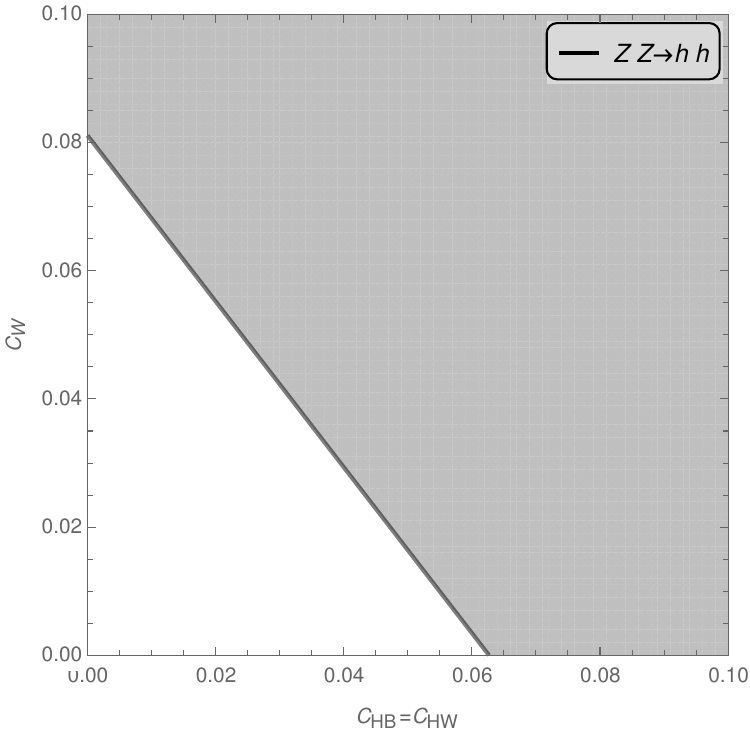}
\end{center} 
\caption{\small The unitarity limits on the WCs, assuming three of the $d=6$ operators are 
generated at the same time, with $C_{HB}$ and $C_{HW}$. Only the 
lower left white portion is allowed.}
\label{fig:double}
\end{figure}

Note that only one operator is taken to be nonzero at a time. 
One may ask whether this is a reasonable assumption, given that almost all the NP 
models necessarily generate more than one effective operators, if not the (almost) full set. Even with 
two such operators, the deviation of the scattering cross-section from the SM expectation can set in 
either before or after the single-operator mark, depending on the signs of the WCs. As a toy example, 
we show, in Fig.\ \ref{fig:double}, the unitarity bound on $\bar c_W$ and $\bar c_{HB}=\bar c_{HW}$,
where all the three operators are present, but the WCs for $O_{HB}$ and $O_{HW}$ are taken to be the 
same for simplicity. Thus, (i) one either 
has to know the ultraviolet complete theory, construct the effective operators, and then study the 
scattering sensitivities, or (ii) perform a complete scanning over the entire 16-dimensional parameter space.
None of them is a viable option. In other words, this study may be useful to find the pattern of nonzero 
WCs if deviations are seen in several channels and are quantified.

Only the worst high-energy behaviour is important; thus, if 
there are terms going as $s^2$ and $s$ in the amplitude, we consider only the $s^2$ term. 
Again, note that the cutoff scale $\Lambda$ has been fixed at 1 TeV just as a fiducial mark and has nothing 
to do with the actual onset of NP. 

In Table \ref{tab:total}, we show which operators affect which $2\to 2$ scattering processes. The 
notation is self-explanatory; the WC, $\bar c_i$, accompanies the operator $O_i$ in Eqs.\ 
(\ref{eqlboson}) and (\ref{eqlfermion}). In Fig.\
\ref{fig:wcbound}, we show the bounds on the corresponding Wilson coefficients of the 
bosonic operators. 
We show only those operators for which one gets an interesting bound that can be probed at the LHC;
thus, $O_6$, $O_H$, and $O_{3W}$ have been dropped, as they do not violate the unitarity bound 
for $\sqrt{s}=2$ TeV. They would do so if we considered $2\to n$ scattering processes but the chances 
of observing them at the LHC is negligible. 
None of the fermionic operators turn out to be interesting; this is also corroborated by Eq.\ (28) of
Ref.\ \cite{1411.5026}.
The corresponding Table \ref{tab:wc} shows the 
point where the Froissart bound is reached. We emphasize again that anomalous behaviour of the 
scattering cross-section should be observable way before this bound is reached. 

 \begin{table}[htbp]
  \centering
   \begin{tabular}{|| c | c | c || c | c | c ||}
    \hline\hline
    WC & Bound & Process & WC & Bound & Process \\ \hline\hline
    $\bar c_W$ & $0.06$     & $WW\to hh$     &  $\bar c_B$ & $0.27$     & $ZZ\to hh$    \\ 
    $\bar c_{HB}$ &  $0.27$    & $ZZ\to hh$     &  $\bar c_\gamma$ &  $0.14$    & $ZZ\to ZZ$     \\ 
    $\bar c_{HW}$ &  $0.06$ & $WW\to hh$ & $\bar c_T$ & $1.2$ & $ZZ\to hh$ \\ \hline
     \hline
   \end{tabular}
   \caption{\small{The limit on the Wilson coefficients, with $\Lambda=1$ TeV. They scale with $\Lambda^2$. 
   Only bounds below $\bar c_i \leq 1.2$ are shown. Gauge boson polarizations are longitudinal. }}
      \label{tab:wc}
 \end{table}
 
One may note that some of the 
coefficients, like $\bar c_{W}$, 
$\bar c_\gamma$, $\bar c_{HB}$, $\bar c_{HW}$, and $\bar c_B$, violate the unitarity bound 
even for relatively smaller values, and increasing $\L$ by a factor of 5 or 10 will still keep them 
in the perturbative domain \footnote{By this, we mean that an ultraviolet-complete theory with perturbative 
couplings can generate such WCs at the low scale. However, we do not consider any possible running 
of these coefficients.}. At the same time, we would like to mention that the anomalous 
behaviour of the scattering amplitudes should be observable much before the unitarity bound is reached, 
and therefore one may surmise the presence of NP for even lower values of the coefficients. Also note that the 
actual coefficient of $O_{3W}$, apart from the loop suppression mentioned before, should be much larger
than that quoted in the Table. 

\begin{figure}[htbp]
\begin{center} 
\includegraphics[width=12cm]{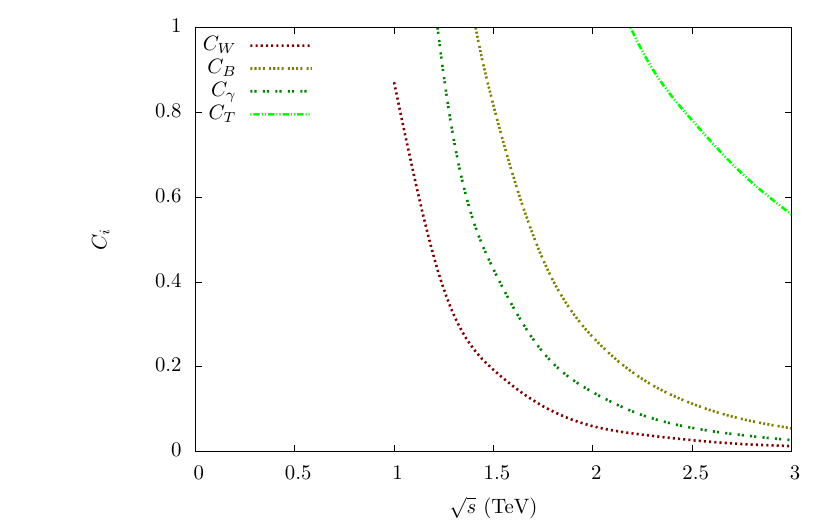}
\end{center} 
\caption{\small The variation of $|C_i|_{\rm max}$, the value where the unitarity limit is reached,
as a function of $\sqrt{s}$. The variation of $C_{HW}$ and $C_{HB}$ are identical to that of 
$C_W$ and $C_B$ respectively.}
\label{fig:ci-var}
\end{figure}

The values of the WCs where unitarity bound is reached decrease with increasing $\sqrt{s}$. 
While our results are displayed for $\sqrt{s}=2$ TeV keeping in mind the maximum partonic centre-of-mass
energy to be had at 14 TeV LHC, we also show how the limits change with $\sqrt{s}$ in Fig.\
\ref{fig:ci-var}. Among the six couplings shown in Table \ref{tab:wc}, 
the variation of $\bar{c}_{HW}$ and $\bar{c}_{HB}$ are identical to that of 
$\bar{c}_W$ and $\bar{c}_B$ respectively.

As we have spent some time on the issue of wavefunction normalization, one may ask how significant 
the corrections are. As can be seen from Table \ref{tab:total} and Fig.\ \ref{fig:wcbound}, such corrections 
can indeed put a strong constraint, {\em e.g.,} on $\bar{c}_B$ and $\bar{c}_\gamma$ from $WW\to WW$ 
scattering. This is because in the amplitude, the field normalization factors appear at the same order of 
$1/\Lambda$ as the vertex corrections.  

There are three points that we would like to mention here. 

1. These bounds are valid if only one operator is present at a time. This may not be the case for the 
particular NP model at hand and there is always the chance of a cancellation, either numerical accident 
or motivated by the theory, in which case the bounds will be strengthened. They can, in principle, also be 
relaxed.  One may, in principle, also get stronger bounds if coupled-channel final states are considered. 

2. The amplitudes start deviating from their SM values much before the $|a_0|=\frac12$ bound is reached. 
Roughly speaking, the amplitudes increase by a factor of 2 over the SM values (which means a fourfold 
increase in the number of final state particles) when the WCs are about one order below their unitarity bound. 
Therefore, a precision measurement can unveil any such new physics much before the unitarity bound 
is reached. In other words, even those operators whose WCs have to be large to hit the unitarity bound (like 
$O_{T}$ or $O_{3W}$) --- or in other words, whose effects are going to be small if the 
couplings are perturbative --- may still be probed in a precision machine. This feature is important if one 
wants to work with perturbative couplings with an increased $\L$. 
At the same time, a detailed study in the moderate-WC region needs the consideration of 
$d > 6$ operators.

3. This is where the future leptonic colliders like the International Linear Collider may have the advantage over 
the LHC, where polarization measurement is going to pose a tough challenge. However, they will lose on the 
$\sqrt{s}$ factor.    

\subsection{Bounds from collider studies}

Effects of the dimension-6 operators in hadronic and leptonic colliders have been studied recently
\cite{1511.05170,1702.05753,ellis,hamzeh} with respect to the SILH Lagrangian \cite{GGPR,contino}, 
using event generators like {\tt MadGraph-aMC@NLO} \cite{madgraph}. The LHC Run-1 data can
provide limits only on the following couplings \footnote{The gluon coupling, $\bar c_g$, is most tightly 
constrained: $\bar c_g \in [-0.01: 0.007]$.}, where we have normalised the mass scale to $\Lambda = 1$
TeV instead of $m_W^2$ or $v^2$ \cite{1511.05170}: 
\be
\bar c_\gamma \in [-0.12 : 0.067]\,,\ \ 
\bar c_{HW} \in [-7.3 : 2.2]\,.
\ee
However, this has been obtained with a fit to the Higgs branching ratios, and the numbers can substantially 
change if the Higgs sector is extended. Thus, while the $\bar c_\gamma$ bound is apparently compatible to the 
unitarity limit, a direct search is always preferred. Such studies have been performed, and the 
reach of the LHC \cite{1702.05753} and the ILC \cite{hamzeh} are as follows:
\bea
\bar c_\gamma &:& [-2.0 : 3.5]~{\rm (LHC300)}\,,\ \ 
[-0.65:1.16]~{\rm (LHC3000)},\ \ [-3.0 : 1.0]~{\rm (ILC)}\,,\nonumber\\
\bar c_{HB} &:& [-5.9 : 7.8]~{\rm (LHC300)}\,,\ \ 
[-1.9 : 2.5] ~{\rm (LHC3000)},\ \ [-2.6 : 1.0]~{\rm (ILC)}\,,\nonumber\\
\bar c_{HW} &:& [-8.2 : 5.9]~{\rm (LHC300)}\,,\ \ 
[-2.6 : 1.9] ~{\rm (LHC3000)},\ \ [-0.29 : 0.27] ~{\rm (ILC)}\,,\nonumber\\
\bar c_W &:& [-0.21 : 0.20]~{\rm (ILC)}\,,\nonumber\\
\bar c_H &:& [-0.65 : 0.67] ~{\rm (ILC)}\,,
\eea
where the LHC numbers are for $\sqrt{s}=14$ TeV with integrated luminosity of 300 fb$^{-1}$ (LHC300)
and 3000 fb$^{-1}$ (LHC3000). The ILC numbers are for $\sqrt{s}=350$ GeV and integrated luminosity of 
3 ab$^{-1}$. Thus, one may hopefully expect some deviation from the SM expectations at the LHC. A study 
on CLIC has also been performed in the second reference of \cite{ellis}, with a slightly different operator basis,
and the sensitivity to the new operators is a bit higher compared to the ILC. 

\section{Summary}

If all the NP fields are heavy (and possibly outside the reach of the LHC), their effects on the SM dynamics
can be parametrized by a set of higher-dimensional operators. These operators spoil the renormalizability 
of the effective theory and in turn can make some scattering amplitudes violate unitarity. Significant 
constraints on the NP parameter space can be obtained if the unitarity violation occurs below the cutoff 
scale $\L$. 

In this paper, we work with a particular basis for the dimension-6 effective operators that are especially 
helpful for scattering studies. However, one can use any such basis for this study, as long as the basis 
is a complete one and does not contain redundant operators. The new operators generate several three-point 
and four-point interactions that contribute to $2\to 2$ scattering processes by modifying the SM vertex factors. 
This in turn spoils the unitarity as bad high-energy behaviours are not cancelled out. 

Some of the new operators also modify the canonical kinetic terms. To bring them back to the canonical forms,
one has to redefine the fields by some multiplicative normalization. Such a redefinition indirectly 
affects the vertices, as can be seen in the list of vertex factors given in Appendix A. Both the effects are important, 
and as can be seen from the plots, a single operator may affect a number of scattering processes,
and a single process may get affected by several operators. We have
followed the approach of minimality
and assumed the presence of only one operators at a time while discussing the bounds. This need not be the 
actual case. 

The bounds depend on the cutoff $\L$ but always scale as $\L^2$, so it is easy to set a fiducial mark at 
$\L = 1$ TeV and show the bounds. They also depend on $\sqrt{s}$ and get stronger as $\sqrt{s}$ increases. 
We have shown all the bounds for $\sqrt{s}=2$ TeV, a typical parton-level energy at the LHC. 

As can be seen, even with $\L=10$ TeV, there are some WCs $\bar c_i$ that remain $\sim {\cal O}(1)$ 
when the unitarity bound $|a_0|=\frac12$ is reached. This is what we can expect very reasonably if the 
NP interaction that generates the effective operators is tree-level and with perturbative couplings. 
At the same time, deviations from SM values can be observed for much smaller values of the WCs. 
However, precise measurement of polarization at the LHC environment remains a challenge to the 
experimentalists.

\centerline{\bf Acknowledgements}

S.G\ thanks the University Grants Commission, Government of India, for a research fellowship. R.I.\ 
and A.K.\ acknowledge the Science and Engineering Research Board, Government of India, for 
support from a research grant. 
A.K.\ also acknowledges the Council for Scientific and Industrial Research, Government of India, for 
another research grant. R.I.\ would also like to thank Sunanda Kumar Patra for help with 
Mathematica. 

\appendix\markboth{Appendix}{Appendix}
\renewcommand{\thesection}{\Alph{section}}
\numberwithin{equation}{section}
\section{Feynman Rules}

We list here all the relevant Feynman rules with bosons in external legs. All momenta are taken to be 
going in to the vertex. The symmetry factors are also included. The SM vertices can be obtained by 
putting all $\bar{c}_i = 0$ or $C_i = 0$, where we use $C_i = \bar{c}_i/\Lambda^2$ for brevity. We also 
use the following shorthand notations already defined in Eq.\ (\ref{field01}), with $e = g s_W = g' c_W$:
\bea
N_h &=& 1 + 2v^2 C_H\,, \nonumber\\
N_Z &=& 1 - \frac12\, {g'}^2 v^2 \left(C_B + 4 s_W^2 C_\gamma\right) -\frac12\, g^2 v^2 C_W \,,\nonumber\\
N_W &=& 1 - \frac12 g^2 v^2 C_W\,,\nonumber\\
N_A &=& 1 - g^2 v^2 s_W^2 C_\gamma\,,\nonumber\\
N_{AZ} &=& \frac14 gg'v^2\left( 8 s_W^2 C_\gamma + C_B - C_W\right)\,.
\eea

We will show here only the unnormalized vertices, {\em i.e.}, vertices obtained with $h,W,Z$ and $A$ and
not their barred (normalized) counterparts. For physical processes, the normalized vertices are relevant. 
To get them, this is what one should do.

\begin{itemize}
\item If the number of external $h$, $W$, and $A$ legs in a vertex be $n_h$, $n_W$, and $n_A$
respectively, divide the vertex factor by $(N_h)^{n_h/2} (N_W)^{n_W/2} (N_A)^{n_A}$. Note the difference 
in the exponent of the photon legs.

\item External $Z$ legs are slightly more complicated. The major contribution comes from the 
unnormalized vertex involving $Z$, and should be divided, in the same vein, by $(N_Z)^{n_Z/2}$ 
where $n_Z$ is the number of $Z$-legs. However, there will also be contributions coming from the 
normalization of the photon fields. For example, suppose we have the following terms in the Lagrangian:
\be
{\cal L} \supset P_{\a\b\mu\nu} W^\a W^\b A^\mu A^\nu + Q_{\a\b\mu\nu} W^\a W^\b A^\mu Z^\nu 
+R_{\a\b\mu\nu}W^\a W^\b Z^\mu Z^\nu
\ee
where $\{P,Q,R\}_{\a\b\mu\nu}$ contain all the momenta and other constants. In terms of the normalized fields,
this becomes
\bea
{\cal L} &\supset& \frac{1}{N_W N_A^2} P_{\a\b\mu\nu} \bar{W}^\a\, \bar{W}^\b\, \bar{A}^\mu\, \bar{A}^\nu 
\nonumber\\
&& + \frac{1}{N_W N_A \sqrt{N_Z}} 
\left( Q_{\a\b\mu\nu} - \frac{2N_{AZ}}{N_A} P_{\a\b\mu\nu} \right) 
\bar{W}^\a\, \bar{W}^\b\, \bar{A}^\mu\, \bar{Z}^\nu\nonumber\\
&& + \frac{1}{N_W N_Z} 
\left( R_{\a\b\mu\nu} - \frac{N_{AZ}}{N_A} Q_{\a\b\mu\nu} + \frac{N_{AZ}^2}{N_A^2} P_{\a\b\mu\nu}\right) 
\bar{W}^\a\, \bar{W}^\b\, \bar{Z}^\mu\, \bar{Z}^\nu\,.
\label{cross}
\eea

\item 
It is trivial to reproduce Table \ref{tab:total} from the vertex factors and relevant field normalizations.
One needs to draw all the possible tree-level diagrams with three- or four-point interactions. If a field 
appears either as an external leg or as an internal propagator in any of these diagrams, the WCs 
involved in its normalization will be affected. For example, If there is any diagram involving $Z$, 
$\bar c_\gamma$, $\bar c_W$, and $\bar c_B$ appear in Table \ref{tab:total}. The WCs coming from 
vertex factors can easily be picked out from the following list.

\end{itemize}  
 Note also that we have never used the on-shell condition $p_i^2 = m_i^2$. If some of the legs are 
on-shell, suitable modifications can be easily performed. The symmetrization over the external momenta has 
also not been done.

\begin{itemize}
 \item Four-scalar vertex:\\
 \be
 h(p_1)h(p_2)h(p_3)h(p_4) : 
  - i \left[ 6\lambda + 45 \lambda v^2 C_6 + 
  4 C_H \sum_{i\not=j} p_i.p_j \right] 
  \ee
  where $i,j$ run from 1 to 4. This has to be divided by $N_h^2$ for the normalized vertex.  
 
 \item Three-scalar vertex:\\
 \be
 h(p_1) h(p_2) h(p_3) : 
 -i v\left[ 6\lambda + 15 \lambda v^2 C_6 + 
 4 C_H \sum_{i\not= j} p_i.p_j\right]
 \ee
 with $i,j = 1,2,3$. This has to be divided by $N_h^{3/2}$ for the normalized vertex. 
 
 \item Four-gauge vertices:\\
 \bea
 W^{+\mu}(p_1) W^{+\nu} (p_2) W^{-\alpha}(p_3) W^{-\beta}(p_4) &:& 
 \frac{ig^2}{4} \left[ - 4 + 4 g^2 v^2 C_W + g^2 v^2 C_{HW}\right]
 \Gamma^{\mu\nu\a\b} + 6ig^4 C_{3W} F^{\mu\nu\a\b}
 \nonumber\\
 %
  W^{+\mu}(p_1) W^{-\nu} (p_2) Z^{\alpha}(p_3) Z^{\beta}(p_4) &:& 
 -\frac{ig^2}{4} \left[ 
 g^2 v^2 C_{HW} + 2 g^2 v^2 C_W (1+c_W^2) - 4 c_W^2\right]
  \Gamma^{\mu\nu\a\b}
 \nonumber\\
 && + 6ig^4 c_W^2 C_{3W} F^{\mu\nu\a\b}\nonumber\\
 %
 W^{+\mu}(p_1) W^{-\nu} (p_2) A^{\alpha}(p_3) Z^{\beta}(p_4) &:& 
 \frac{ig^2} {8}
  \left[g g' v^2 C_{HW} + 2 g g' v^2 C_W (1+2 c_W^2) 
 -s_W c_W\right]
\Gamma^{\mu\nu\a\b}\nonumber\\
&& - 6ig^4 s_W c_W C_{3W} F^{\mu\nu\a\b}\,,\nonumber\\
 W^{+\mu}(p_1) W^{-\nu} (p_2) A^{\alpha}(p_3) A^{\beta}(p_4) &:& 
 \frac{ie^2}{2} \left(2-g^2 v^2 C_W\right)  \Gamma^{\mu\nu\a\b}
 \nonumber\\
  && - 6ig^4 s_W^2 C_{3W} F^{\mu\nu\a\b}
 \,,
 \eea
 where
 \be
 \Gamma^{\mu\nu\a\b} = \left(\eta^{\mu\b}\eta^{\nu\a} + \eta^{\mu\a}\eta^{\nu\b} 
 -2 \eta^{\mu\nu} \eta^{\a\b}\right)\,,
 \ee
 and
 \bea
 F^{\mu\nu\a\b}&=& 
 (p_1.p_3+p_2.p_4)\eta^{\mu\b}\eta^{\nu\a} + (p_1.p_4+p_2.p_3)\eta^{\mu\a}\eta^{\nu\b} 
 - (p_1+p_2).(p_3+p_4)\eta^{\mu\nu}\eta^{\a\b}\nonumber\\
 && + \eta^{\mu\nu}((p_1+p_2)^\b p_4^\a + (p_1+p_2)^\b p_3^\a) 
 + \eta^{\a\b} ((p_3+p_4)^\mu p_1^\nu + (p_3+p_4)^\nu p_2^\mu) \nonumber\\
 && + \eta^{\mu\a} (p_1^\b(p_3-p_4)^\nu - p_2^\b p_3^\nu - p_1^\nu p_3^\b)
  +\eta^{\nu\b} (p_4^\mu (p_2-p_1)^\a -p_2^\a p_3^\mu - p_2^\mu p_4^\a) \nonumber\\
  && + \eta^{\mu\b}(p_1^\a (p_4-p_3)^\nu - p_2^\a p_4^\nu  - p_1^\nu p_4^\a) 
  +\eta^{\nu\a}(p_3^\mu(p_2-p_1)^\b - p_2^\b p_4^\mu - p_2^\mu p_3^\b)\,.
 \eea
  Note that the $WW\gamma\gamma$ four-point vertex 
  contributes to $WWZZ$ and $WWAZ$ vertices as shown in Eq.\ (\ref{cross}). 
 
 \item Three-gauge vertices:\\
 \bea
 W^{+\mu}(p_1) W^{-\nu} (p_2) Z^{\alpha}(p_3) &:& 
 \frac{ig}{8c_W} \left[ 
 -8 c_W^2 \left( (p_1-p_2)^\a\eta^{\mu\nu} +(p_2-p_3)^\mu\eta^{\nu\a} + 
 (p_3-p_1)^\nu\eta^{\mu\a} \right)\right.\nonumber\\
 && \left. + 48 g^2 c_W^2 C_{3W} \left[ (p_1^\nu p_2^\a p_3^\mu - p_1^\a p_2^\mu p_3^\nu) 
 + p_2.p_3(-p_1^\nu \eta^{\a\mu} + p_1^\a \eta^{\mu\nu}) \right. \right.\nonumber\\
 && \left. \left. 
 + p_1.p_3(-p_2^\a \eta^{\mu\nu} + p_2^\mu \eta^{\nu\a})
 + p_1.p_2(-p_3^\mu \eta^{\nu\a} + p_3^\nu \eta^{\mu\a})\right]\right.\nonumber\\  
  && \left. + 2g^2 v^2 C_W \left(
 (1+2c_W^2) (p_1-p_2)^\a\eta^{\mu\nu} + s_W^2(p_1^\mu\eta^{\nu\a}- p_2^\nu\eta^{\mu\a}) 
 \right.\right.\nonumber\\
 && \left.\left. - (2+c_W^2) (p_1^\nu\eta^{\mu\a} - p_2^\mu\eta^{\nu\a})
 + 3c_W^2(p_3^\nu\eta^{\mu\alpha}-p_3^\mu\eta^{\nu\alpha})
 \right)\right.\nonumber\\
 &&\left. + g^2 v^2 C_{HW} \left(
 (p_1-p_2)^\a\eta^{\mu\nu} +  p_2^\mu\eta^{\nu\a} - p_1^\nu\eta^{\mu\a}
 +c_W^2 (p_3^\nu\eta^{\mu\alpha}-p_3^\mu\eta^{\nu\alpha}) 
 \right)\right.\nonumber\\
 &&\left. -g^2 v^2 s_W^2 C_{HB}  (p_3^\nu\eta^{\mu\alpha}-p_3^\mu\eta^{\nu\alpha})   \right]\,,\nonumber\\ 
 A^\mu(p_1) W^{+\nu}(p_2) W^{-\a} (p_3) &:& 
 \frac{ie}{8} \left[-8 \left(
 (p_1-p_2)^\a \eta^{\mu\nu} + (p_2-p_3)^\mu \eta^{\nu\a} + (p_3-p_1)^\nu \eta^{\mu\a}\right)\right.
 \nonumber\\
 && \left. + 48 g^2 C_{3W} \left[ (p_1^\nu p_2^\a p_3^\mu - p_1^\a p_2^\mu p_3^\nu) 
 + p_2.p_3(-p_1^\nu \eta^{\a\mu} + p_1^\a \eta^{\mu\nu}) \right. \right.\nonumber\\
 && \left. \left. 
 + p_1.p_3(-p_2^\a \eta^{\mu\nu} + p_2^\mu \eta^{\nu\a})
 + p_1.p_2(-p_3^\mu \eta^{\nu\a} + p_3^\nu \eta^{\mu\a})\right]\right.\nonumber\\ 
  && \left. + 2g^2v^2 C_W\left( 3(p_1^\a\eta^{\mu\nu}-p_1^\nu\eta^{\mu\a}) -(p_2-p_3)^\a
 \eta^{\mu\nu} - (p_2-p_3)^\nu \eta^{\mu\a} \right.\right.\nonumber\\
 &&\left.\left. + 2(p_2-p_3)^\mu\eta^{\nu\a} \right)
+ g^2 v^2 (C_{HW}+C_{HB}) (p_1^\a \eta^{\mu\nu} - p_1^\nu\eta^{\mu\a}) \right] 
 \eea
 
 \item Three-point mixed vertices:\\
 \bea
 h(p_1) A^\mu(p_2) Z^\nu(p_3) &:& -\frac{igg' v}{4} \, \left[
 -2 (p_2^\mu p_2^\nu - p_2^2 \eta^{\mu\nu}) (C_B-C_W) 
 \right. \nonumber\\
 && \left.  +  (p_1^\mu p_2^\nu - p_1.p_2 \eta^{\mu\nu}) (C_{HB}-C_{HW})
 +  16 s_W^2 C_\gamma (p_3^\mu p_2^\nu - \eta^{\mu\nu}p_2.p_3)
 \right]\,,\nonumber\\
 h(p_1) W^{+\mu}(p_2) W^{-\nu}(p_3) &:& 
 \frac{ig^2v}{4}  \left[ 2\eta^{\mu\nu} 
 +2 C_W\left( p_2^2\eta^{\mu\nu}-p_2^\mu p_2^\nu + p_3^2\eta^{\mu\nu}-p_3^\mu p_3^\nu\right)\right.
 \nonumber\\
 && \left. + C_{HW} \left( p_1^\mu p_2^\nu + p_1^\nu p_3^\mu - (p_1.(p_2+p_3)\eta^{\mu\nu}\right) 
 \right]\,,\nonumber\\
 h(p_1) Z^{\mu}(p_2) Z^{\nu}(p_3) &:& 
 \frac{ig^2 v}{4c_W^2} \left[ 2\eta^{\mu\nu} + 2 (c_W^2 C_W
 +s_W^2 C_B) \left(p_3^2\eta^{\mu\nu} 
 -p_3^\mu p_3^\nu + p_2^2\eta^{\mu\nu} - p_2^\mu p_2^\nu\right) \right.\nonumber\\
 && \left. + 16 s_W^4 C_\gamma \left(p_3^\mu p_2^\nu - p_2.p_3 \eta^{\mu\nu}\right) 
 - 8 v^2 C_T \eta^{\mu\nu} \right. \nonumber\\
 && \left. \left( c_W^2 C_{HW} + s_W^2 C_{HB} \right) \left( 
 p_1^\mu p_2^\nu - p_1.p_2 \eta^{\mu\nu} + p_3^\mu p_1^\nu - p_1.p_3 \eta^{\mu\nu} 
 \right) \right]\,, \nonumber\\
 h(p_1) A^\mu(p_2) A^\nu (p_3) &:& 4ie^2 v C_\gamma
 \left(p_1^\nu p_2^\mu - p_1.p_2 \eta^{\mu\nu}\right)\,,
 \eea
 
 Note that the $hAZ$ and $hAA$ vertices are generated by the new operators only. 
 
 \item Four-point mixed vertices:\\
 \bea
 h(p_1) h(p_2) W^{+\mu}(p_3) W^{-\nu}(p_4) &:& 
 \frac{ig^2}{4} \left[ 
 2\eta^{\mu\nu} + 2C_W\left( p_3^2\eta^{\mu\nu} - p_3^\mu p_3^\nu 
 + p_4^2\eta^{\mu\nu}-p_4^\mu p_4^\nu\right)\right.\nonumber\\
 && \left. + C_{HW} \left( 
 (p_1+p_2)^\mu p_3^\nu 
 +(p_1+p_2)^\nu p_4^\mu -(p_1+p_2).(p_3+p_4) \eta^{\mu\nu} \right)\right]\,,
 \nonumber\\
 h(p_1) h(p_2) Z^{\mu}(p_3) Z^{\nu}(p_4) &:& 
 \frac{ig^2}{4c_W^2} \left[2\eta^{\mu\nu} - 24 v^2 C_T \eta^{\mu\nu} +16 s_W^4 C_\gamma
 (p_3^\nu p_4^\mu-p_3.p_4\eta^{\mu\nu})\right.\nonumber\\
 &&\left. - 2 c_W^2 C_W\left( p_3^\mu p_3^\nu - p_3^2 \eta^{\mu\nu} + 
 p_4^\mu p_4^\nu - p_4^2 \eta^{\mu\nu}\right) \right. \nonumber\\
  &&\left. - 2 s_W^2 C_B\left( p_3^\mu p_3^\nu - p_3^2 \eta^{\mu\nu} + 
 p_4^\mu p_4^\nu - p_4^2 \eta^{\mu\nu}\right) \right. \nonumber\\ 
 && \left. + \left( c_W^2 C_{HW} + s_W^2 C_{HB}\right) \times\right.\nonumber\\ 
 && \left. \left(
 (p_1+p_2)^\mu p_3^\nu + (p_1+p_2)^\nu p_4^\mu - (p_1+p_2).(p_3+p_4)\eta^{\mu\nu} \right)
 \right] \,,\nonumber
 \eea
 \bea
   h(p_1) W^{+\mu}(p_2) W^{-\nu}(p_3) Z^\a (p_4) &:& 
 \frac{ig^3v}{4c_W} 
 \left[ C_W\left( (2+4c_W^2) (p_2-p_3)^\a\eta^{\mu\nu} \right.\right.\nonumber\\
 && \left. \left. + (6 p^\nu_4c_W^2 - 2 p^\nu_3 s_W^2 - p_2^\nu(4+2c_W^2) \eta^{\mu\a} \right.\right.\nonumber\\
 &&\left.\left. 
 - (6 p^\mu_4c_W^2 - 2 p^\mu_2 s_W^2 - p_3^\mu(4+2c_W^2) \eta^{\nu\a}
 \right)\right.\nonumber\\
 && \left. 
 + C_{HW} \left((p_2-p_3)^\a\eta^{\mu\nu} + (p_4^\nu c_W^2 + p_1^\nu s_W^2 -p_2^\nu)\eta^{\mu\a}
 \right.\right.\nonumber\\
 &&
 \left.\left. -(p_4^\mu c_W^2 + p_1^\mu s_W^2 -p_3^\mu)\eta^{\nu\a} 
 \right) \right.\nonumber\\
 &&
 \left. +C_{HB} s_W^2 \left( p_4^\mu\eta^{\nu\a}-p_4^\nu\eta^{\mu\a}\right)\right]\,,\nonumber\\
 h(p_1) A^\mu(p_2) W^{+\nu}(p_3) W^{-\a}(p_4) &:& 
 \frac{i e g^2 v}{4} \left[ 
 C_W \left( 6(p_2^\a\eta^{\mu\nu}-p_2^\nu\eta^{\mu\a}) \right.\right.\nonumber\\
 && \left.\left. -2[(p_3-p_4)^\a\eta^{\mu\nu} + (p_3-p_4)^\nu\eta^{\mu\a} -2(p_3-p_4)^\mu
 \eta^{\nu\a}]\right)\right.\nonumber\\
 &&\left. + C_{HW} \left( (p_1-p_2)^\nu\eta^{\mu\a}-(p_1-p_2)^\a\eta^{\mu\nu}\right) \right.\nonumber\\
 && \left. + C_{HB} (p_2^\a\eta^{\mu\nu}-p_2^\nu \eta^{\mu\a}) \right]\,,\nonumber\\
 h(p_1)h(p_2)A^\mu(p_3)A^\nu(p_4) &:& 
 4ie^2 C_\gamma (p_3^\nu p_4^\mu - p_3.p_4 \eta^{\mu\nu})\,,\nonumber\\
 h(p_1)h(p_2)A^{\mu}(p_3)Z^\nu(p_4) &:& 
 -\frac{igg'}{4}\left[
 2 (C_B-C_W) (p_3^2\eta^{\mu\nu}-p_3^\mu p_3^\nu)\right.\nonumber\\
 && \left.  + 16 s_W^2 C_\gamma
 (p_3^\nu p_4^\mu - p_3.p_4 \eta^{\mu\nu} ) \right. \nonumber\\
 && \left. +(C_{HB}-C_{HW}) (p_3^\nu(p_1+p_2)^\mu - p_3.(p_1+p_2)\eta^{\mu\nu})\right]
 \eea
 
 Last four vertices have been generated only through the effective operators. 
 
 \item Three-point fermionic vertices:
 \bea
 \bar b (p_1) b (p_2) h (p_3)
&:&
- \frac{im_b}{v} \left[ 1 + \frac{3}{2} C_d v^2 \right]\,,\nonumber
\\
\bar t (p_1) t (p_2) h (p_3)
&:&
- \frac{im_t}{v} \left[ 1 + \frac{3}{2} C_u v^2 \right]\,,\nonumber\\
\bar b (p_1) b (p_2) A^\mu (p_3)
&:&
- \frac{ie}{3} \gamma^\mu
+e \left( C_{dB} - C_{dW}\right)  m_b \sigma^{\mu\nu} p_{3\nu}\,,\nonumber
\\
\bar t (p_1) t (p_2) A^\mu (p_3)
&:&
\frac{2ie}{3} \gamma^\mu
+e \left( C_{uB} + C_{uW}\right) m_t \sigma^{\mu\nu} p_{3\nu}\,,\nonumber
\\
\bar b (p_1) b (p_2) Z^\mu (p_3)
&:&
  \frac{ig}{2c_W} \gamma^\mu \left[ \frac{2}{3} s^2_W - P_L 
  - \left( C_{Hq} + C'_{Hq}\right) v^2 P_L
                         - C_{Hd} v^2 P_R  \right]\nonumber\\
&&
- \frac{g}{c_W} \left[ \left(C_{dB} s^2_W + C_{dW} c^2_W\right) m_b \sigma^{\mu\nu} p_{3\nu}
                           \right]\,,\nonumber
\\
\bar t (p_1) t (p_2) Z^\mu (p_3)
&:&
  \frac{ig}{2c_W} \gamma^\mu \left[ P_L - \frac{4}{3} s^2_W 
  - \left(C_{Hq} - C'_{Hq}\right) v^2 P_L
                         - C_{Hu} v^2  P_R  
  \right]\nonumber\\
&&
- \frac{g}{c_W} \left[
                         \left( C_{uB} s^2_W - C_{uW} c^2_W\right) m_t \sigma^{\mu\nu} p_{3\nu}
                           \right]\,,\nonumber\\
\bar t (p_1) b (p_2) W^{-\mu} (p_3)
&:&
\frac{ig}{\sqrt{2}} V_{tb} \left[ 1 + C'_{Hq} v^2 \right] \gamma^\mu P_L
+ \frac{ig}{\sqrt{2}} C_{Hud} v^2 \gamma^\mu P_R \nonumber\\
&&
+ \sqrt{2}\, g V_{tb} \left[
                         m_t C_{uW} \sigma^{\mu\nu} p_{3\nu} P_L
                         +m_b C_{dW}\sigma^{\mu\nu} p_{3\nu} P_R
                           \right]\,.
\eea

\item Four-point fermionic vertices
\bea
\bar b (p_1) b (p_2) W^{-\mu} (p_3) W^{+\nu} (p_4)
&:&
g^2 \, m_b C_{dW} \sigma^{\mu\nu}\,,\nonumber
\\
\bar t (p_1) t (p_2) W^{-\mu} (p_3) W^{+\nu} (p_4)
&:&
  -g^2\, m_t C_{uW} \sigma^{\mu\nu}\,,\nonumber
\\
\bar t (p_1) b (p_2) W^{-\mu} (p_3) Z^\nu (p_4)
&:&
\sqrt{2}\, g^2 c_W V_{tb} \left[ m_t C_{uW} \sigma^{\mu\nu} P_L + 
m_b C_{dW} \sigma^{\mu\nu} P_R \right]\,.
\eea

\end{itemize}

There is no field normalization for the fermions as their kinetic terms are not affected but the bosonic 
fields need to be normalized, and as before, $t\bar{t}(b\bar{b})Z$ vertices get a contribution from 
$t\bar{t} (b\bar{b}) A$.

\section{The operator in detail}

It is instructive to write out at least one of the operators in detail to show the 
rich structure. For example, the operator $O_W \equiv 
\left(\Phi^\dag \tau^i \overleftrightarrow D^\mu \Phi\right)\left(D^\nu W_{\mu\nu} \right)^i$
looks like (note that $\bar c_W$ is the WC while $c_W$ is the cosine of the Weinberg angle):

\bea
 O_W
 &=&
 \frac{g^2}{4} \left[ v^2
 \left[
   (\p^\mu W^{+\nu} - \p^\nu W^{+\mu}) (\p_\mu W^-_\nu - \p_\nu W^-_\mu)
 \right.\right.
 \nonumber\\
 &&
  + i g c_W (W^{+\mu} \p^\nu W^-_\nu - W^{-\mu} \p^\nu W^+_\nu) Z_\mu
  - i g c_W ( W^{+\mu} W^{-\nu} - W^{-\mu} W^{+\nu} ) (\p_\mu Z_\nu - 2 \p_\nu Z_\mu)
 \nonumber\\
 &&
  + i g c_W W^{+\mu} (\p_\mu W^-_\nu - 2 \p_\nu W^-_\mu) Z^\nu
  - i g c_W W^{-\mu} (\p_\mu W^+_\nu - 2 \p_\nu W^+_\mu) Z^\nu
 \nonumber\\
 &&
  + i g s_W (W^{+\mu} \p^\nu W^-_\nu - W^{-\mu} \p^\nu W^+_\nu) A_\mu
  - i g s_W ( W^{+\mu} W^{-\nu} - W^{-\mu} W^{+\nu} ) (\p_\mu A_\nu - 2 \p_\nu A_\mu)
 \nonumber\\
 &&
  + i g s_W W^{+\mu} (\p_\mu W^-_\nu - 2 \p_\nu W^-_\mu) A^\nu
  - i g s_W W^{-\mu} (\p_\mu W^+_\nu - 2 \p_\nu W^+_\mu) A^\nu
 \nonumber\\
 &&
  - g^2 (W^{+\mu} W^{-\nu} - W^{+\nu} W^{-\mu}) (W^+_\mu W^-_\nu - W^+_\nu W^-_\mu)
 \nonumber\\
 &&
  - g^2 c^2_W (W^{+\mu} W^{-\nu} + W^{-\mu} W^{+\nu}) Z_\mu Z_\nu
  + 2 g^2 c^2_W W^{-\mu} W^+_\mu Z^\nu Z_\nu
 \nonumber\\
 &&
  - g^2 c_W s_W (W^{+\mu} W^{-\nu} + W^{-\mu} W^{+\nu}) (Z_\mu A_\nu + A_\mu Z_\nu)
  + 4 g^2 c_W s_W W^{-\mu} W^+_\mu Z^\nu A_\nu
 \nonumber\\
 &&
 \left.
  - g^2 s^2_W (W^{+\mu} W^{-\nu} + W^{-\mu} W^{+\nu}) A_\mu A_\nu
  + 2 g^2 s^2_W W^{-\mu} W^+_\mu A^\nu A_\nu
 \right]
 \nonumber\\
 &&
 + \frac{v^2}{c_W}
 \left[
  - c_W \p^\nu Z^\mu (\p_\mu Z_\nu - \p_\nu Z_\mu)
  - s_W \p^\nu Z^\mu (\p_\mu A_\nu - \p_\nu A_\mu)
 \right.
 \nonumber\\
 &&
  - i g Z^\mu (W^+_\mu \p^\nu W^-_\nu - W^-_\mu \p^\nu W^+_\nu)
 \nonumber\\
 &&
  - i g Z^\mu W^{+\nu} (\p_\mu W^-_\nu - 2 \p_\nu W^-_\mu)
  + i g Z^\mu W^{-\nu} (\p_\mu W^+_\nu - 2 \p_\nu W^+_\mu)
 \nonumber\\
 &&
  - g^2 c_W (W^{+\mu} W^{-\nu} + W^{-\mu} W^{+\nu}) Z_\mu Z_\nu
  + 2 g^2 c_W W^{+\nu} W^-_\nu Z^\mu Z_\mu
 \nonumber\\
 &&
 \left.
  - g^2 s_W (W^{+\mu} W^{-\nu} + W^{-\mu} W^{+\nu}) Z_\mu A_\nu
  + 2 g^2 s_W W^{+\nu} W^-_\nu Z^\mu A_\mu
 \right]
 \nonumber\\
 &&
 + 2vh
 \left[
   (\p^\mu W^{+\nu} - \p^\nu W^{+\mu}) (\p_\mu W^-_\nu - \p_\nu W^-_\mu)
 \right.
 \nonumber\\
 &&
  + i g c_W (W^{+\mu} \p^\nu W^-_\nu - W^{-\mu} \p^\nu W^+_\nu) Z_\mu
  - i g c_W ( W^{+\mu} W^{-\nu} - W^{-\mu} W^{+\nu} ) (\p_\mu Z_\nu - 2 \p_\nu Z_\mu)
 \nonumber\\
 &&
  + i g c_W W^{+\mu} (\p_\mu W^-_\nu - 2 \p_\nu W^-_\mu) Z^\nu
  - i g c_W W^{-\mu} (\p_\mu W^+_\nu - 2 \p_\nu W^+_\mu) Z^\nu
 \nonumber\\
 &&
  + i g s_W (W^{+\mu} \p^\nu W^-_\nu - W^{-\mu} \p^\nu W^+_\nu) A_\mu
  - i g s_W ( W^{+\mu} W^{-\nu} - W^{-\mu} W^{+\nu} ) (\p_\mu A_\nu - 2 \p_\nu A_\mu)
 \nonumber\\
 &&
 \left.
  + i g s_W W^{+\mu} (\p_\mu W^-_\nu - 2 \p_\nu W^-_\mu) A^\nu
  - i g s_W W^{-\mu} (\p_\mu W^+_\nu - 2 \p_\nu W^+_\mu) A^\nu
 \right]
 \nonumber\\
 &&
 + \frac{2vh}{c_W}
 \left[
  - c_W \p^\nu Z^\mu (\p_\mu Z_\nu - \p_\nu Z_\mu)
  - s_W \p^\nu Z^\mu (\p_\mu A_\nu - \p_\nu A_\mu)
 \right.
 \nonumber\\
 &&
  - i g Z^\mu (W^+_\mu \p^\nu W^-_\nu - W^-_\mu \p^\nu W^+_\nu)
 \nonumber\\
 &&
 \left.
  - i g Z^\mu W^{+\nu} (\p_\mu W^-_\nu - 2 \p_\nu W^-_\mu)
  + i g Z^\mu W^{-\nu} (\p_\mu W^+_\nu - 2 \p_\nu W^+_\mu)
 \right]
 \nonumber\\
 &&
 + h^2
   (\p^\mu W^{+\nu} - \p^\nu W^{+\mu}) (\p_\mu W^-_\nu - \p_\nu W^-_\mu)
 \nonumber\\
 &&
 \left.
 - h^2
   \p^\nu Z^\mu (\p_\mu Z_\nu - \p_\nu Z_\mu)
 - \frac{s_W}{c_W} h^2
   \p^\nu Z^\mu (\p_\mu A_\nu - \p_\nu A_\mu)\right]
 \label{OW3}
\eea



\end{document}